\documentclass[aps,prl,superscriptaddress,amsmath,twocolumn,showpacs]{revtex4}
\usepackage{epsfig}
\usepackage{float}

\def\ai         {\emph{ab initio}\;}
\def\ga         {\alpha}

\def\gd         {\delta} 
\def\eps        {\epsilon}
\def\gee        {\varepsilon}
\def\gl         {\lambda}
\def\go         {\omega}

\def\qlim       {\qq\rightarrow{\bf 0}}

\def\la         {\langle}
\def\ra         {\rangle}

\def\qq         {{\bf q}}

\renewcommand{\[}{\left[}
\renewcommand{\]}{\right]}
\renewcommand{\(}{\left(}
\renewcommand{\)}{\right)}

\restylefloat{figure}

\begin{document}
\title{
Exciton--plasmon states in nanoscale materials: breakdown
of the Tamm--Dancoff approximation}
\author{Myrta Gr\"uning}
\affiliation{European Theoretical Spectroscopy Facility (ETSF),
Universit\'e Catholique de Louvain, 
Unit\'e de Physico-Chimie et de Physique des Mat\'eriaux, B-1348
Louvain-la-Neuve, Belgium}
\author{Andrea Marini}
\affiliation{
European Theoretical Spectroscopy Facility (ETSF),
CNR-INFM Institute for Statistical Mechanics and Complexity,
CNISM and Dipartimento di Fisica, Universit\'a di Roma ``Tor Vergata'',
via della Ricerca Scientifica 1, 00133 Roma, Italy}
\author{Xavier Gonze}
\affiliation{European Theoretical Spectroscopy Facility (ETSF),
Universit\'e Catholique de Louvain, 
Unit\'e de Physico-Chimie et de Physique des Mat\'eriaux, B-1348
Louvain-la-Neuve, Belgium}

\begin{abstract}
Within the Tamm--Dancoff approximation
\ai approaches describe excitons 
as packets of electron-hole pairs propagating only
forward in time. 
However, we show that 
in nanoscale materials 
excitons and plasmons hybridize, creating exciton--plasmon states
where the electron-hole pairs oscillate
back and forth in time. Then, as exemplified by 
the {\it trans}-azobenzene molecule and carbon nanotubes,
the Tamm--Dancoff approximation yields errors as 
large as the accuracy claimed in \ai calculations. 
Instead, we propose a general and efficient approach that
avoids the Tamm--Dancoff approximation, and correctly describes
excitons, plasmons and exciton--plasmon states.
\end{abstract}
%
%
%
%
%
\maketitle

The Bethe--Salpeter\,(BS)~\cite{SalpeterB51}  and 
the Time-Dependent Density Functional Theory\,(TDDFT)~\cite{RungeG84} 
equations
allow the accurate
calculation of the polarization function of many physical systems  without relying
on external parameters. Within these frameworks neutral
excitations are described as combination of electron-hole\,(e-h)
pairs of a noninteracting system. 
However for nanoscale materials, 
the huge number of e-h
pairs involved makes the solution of the BS/TDDFT 
equation extremely cumbersome. Consequently, the increasing interest in the excitation 
properties of such materials has justified the use of
\emph{ad-hoc} approximations. 
The most important and widely-used
is the  Tamm-Dancoff approximation\,(TDA)~\cite{FetterW71} where only positive energy 
e-h pairs are considered. Within the TDA the interaction between
e-h pairs at positive and negative (antipairs) energies is neglected, and only one e-h pair is assumed to 
propagate in any time interval.
The main advantage of the TDA is that 
the \emph{non-Hermitian} BS/TDDFT problem reduces
to a \emph{Hermitian} problem, that can be solved with efficient and
stable iterative methods~\cite{Bai00}.

In Solid State Physics---a major field of
application of the BS/TDDFT equation---the success of TDA is based on the sharp distinction between
excitonic and plasmonic excitations. 
Excitons are {\it localized} packets of e-h pairs
bound together by the Coulomb attraction and are observed in 
optical absorption experiments. Plasmons are, instead, {\it delocalized} collective 
oscillations of the electronic density that induce a
macroscopic polarization effect and are observed 
in electron energy loss\,(EEL)
experiments.  In contrast to the case of excitons,
the TDA is known to misdescribe plasmons in solids~\cite{OlevanoR01} as
the density oscillations involve the excitation of e-h antipairs.
Nevertheless, the success in describing optical absorption of 
solids and the remarkable numerical advantages have motivated  
the application of the TDA
to very different systems.  
Nowadays the BS/TDDFT equation \emph{within the TDA} is
becoming a standard tool to study excitations in
nanostructures~\cite{More,Spataru08}, and in
molecular systems~\cite{Hirata99}.

In this Letter we argue that for confined systems---such
as nanostructures or $\pi$-conjugated molecules---the excitations appearing in the
response function show a mixed excitonic--plasmonic behavior.
As a consequence the e-h pair-antipair interaction
becomes crucial and the TDA does not hold anymore. 
A paradigmatic example  is the {\it trans}-azobenzene  molecule,
where the TDA overestimates the static polarizability  by $\sim$40\%.
This error is larger than the claimed accuracy in \ai calculations. 
Even more intriguing is the case of
carbon nanotubes that, because of
the quasi-one-dimensional (1D) structure,
behave either as extended or isolated system depending on the
direction of the perturbing field.
Thus, for transverse perturbations the excitons acquire a plasmonic nature and the
TDA overestimates the position of the $\pi$ plasmon peak appearing in both absorption
and EEL spectra by almost 1\,eV.
By exploiting the symmetry properties of the BS and TDDFT kernels 
we devise a robust and efficient iterative approach to calculate the
frequency-dependent response,
beyond the TDA. This approach benefits from the same numerical advantages of
Hermitian techniques,  and 
correctly describes excitons, plasmons 
and exciton--plasmon states. 

To introduce and understand the reasons beyond the breakdown of the TDA we need to 
study in detail the structure of the TDDFT and BS equations.
These are commonly rewritten as a
Hamiltonian problem~\cite{OnidaRR02}, by expanding the single particle states in the Kohn--Sham 
basis. 
Then, the BS/TDDFT Hamiltonian $H$ is a matrix in the Fock space of the e-h pairs $|eh\ra$ and
antipairs $|\widetilde{he}\ra$, and
it has the block-form 
\begin{equation}\label{eq:effham}
H =
\begin{pmatrix}
R  & C \\
-C^{*}  & -R^*
\end{pmatrix}.
\end{equation}
The resonant block $R$ is Hermitian 
and the coupling
block $C$ is symmetric (see Appendix B of Ref.~\cite{OnidaRR02}). 
The dielectric function $\gee\(\go\)$ is written in terms of the
resolvent of $H$, $\(\go-H\)^{-1}$, as
$\gee\(\go\)=1-\(8\pi\)/\Omega\la P | \(\go-H+i0^+\)^{-1}|P\ra$, 
where $\Omega$ is the simulation volume.
In the limit of large $\Omega$ 
the polarizability is given by $\ga\(\go\)\propto \gee\(\go\)$. 
$|P \rangle$ is a ket whose components along the $|eh\ra$
space are the optical oscillators:
$\la P | eh\ra \sim \la e | \vec{d}\cdot\vec\xi|h \ra$, with 
$\vec{d}$ the electronic dipole, and $\vec\xi$ the light polarization 
factor.

\begin{figure}[H]
\begin{center}
\epsfig{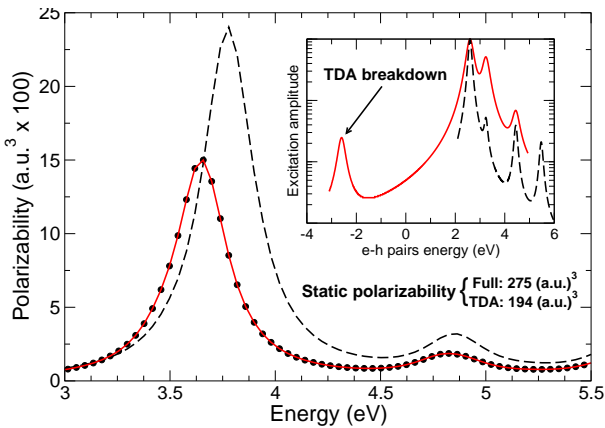}
\end{center}
\vspace{-.5cm}
\caption{
\footnotesize{
Dynamical polarizability of the \emph{trans}-azobenzene molecule calculated within
the TDLDA either by using the full Hamiltonian (solid
line) or the TDA (dashed line). For comparison the results obtained
by diagonalization (circles) are also reported. 
The TDA largely underestimates the static polarizability, by almost 40\,\%.
More importantly the amplitude function (inset)
of the most intense peak of the polarizability (see text) differs
dramatically in the TDA and in the full Hamiltonian. 
The TDA misses of an important contribution from  
the spatially extended e-h antipair with energy $\sim-2.7$\,eV.
}}
\label{fg:azospc}
\end{figure}

The matrix elements of $R$ and $C$ have different definitions in the BS and TDDFT
Hamiltonians. Within the BS equation
$R_{\substack{ee'\\hh'}}=E_{eh}\delta_{ee'}\delta_{hh'} + \la e h| W-2\bar{V}|e'h'\ra$, and
$C_{\substack{ee'\\hh'}}=\la e h| W-2\bar{V}|\widetilde{h'e'}\ra$. $E_{eh}$ is the
energy of the independent e-h pair,  
$W$ is the statically screened e-h attraction and $\bar{V}$ is the 
bare Coulomb interaction without the long-range tail~\cite{OnidaRR02}.
Within TDDFT  $W$ is replaced by $-2f_{\text{xc}}$, with 
$f_{\text{xc}}$ the exchange--correlation kernel. 

The TDA assumes that the
effect induced by the e-h antipairs is negligible.
Consequently, the $C$ block, that couples pairs and antipairs, 
is neglected, and the Hamiltonian $H$
is approximated by $R$.  
The coupling described by $C$ is dominated by the contribution arising from the
bare Coulomb interaction $\bar{V}$.
This term, called Hartree contribution, 
measures the degree of inhomogeneity of the electronic density. 
The larger this inhomogeneity,
the stronger the corresponding polarization and the induced local
fields that 
counteract the external perturbing field.
Therefore, the inhomogeneity of the electronic density and the
strength of local fields discriminate whether an excitation is well-described by the TDA.
Furthermore the TDA is known to fail for plasmonic
excitations~\cite{OlevanoR01} that, causing an oscillation of  the density, 
involve the creation of e-h pairs and antipairs.

The electronic density of confined systems, like molecules and nanostructures, is typically strongly inhomogeneous.  
Moreover the excitons can be spread all over the molecule,
involving the excitations of most of the electrons. Thus, in contrast to solids,
it is not possible
to distinguish
between excitonic and plasmonic excitations, and
the arguments commonly used to sustain the TDA fail.

Indeed, the striking failure of the TDA is clearly demonstrated by 
the dynamical
polarizability $\Im\[\ga\(\go\)\]$ of
\emph{trans}-azobenzene,
calculated within the time-dependent local density approximation (TDLDA)~\cite{note1,note2}.
In  Fig.~\ref{fg:azospc}
we compare $\Im\[\ga\(\go\)\]$ 
calculated either using the TDA, or by solving the full $H$ eigenproblem.
TDA yields a blueshift of 0.2\,eV and
a large overestimation of the intensity of the main peak. 
More importantly the TDA  causes a 
40\% underestimation of the static polarizability, $\Re\[\ga\(\go=0\)\]$.
The reason for this failure can be understood looking at the
amplitude function $A^{\gl}\(\go\)$ of the eigenstates $|\gl\ra$ of $H$, defined as
$A^{\gl}\(\go\)= \sum_{\eta=\{eh\},\{\widetilde{eh}\}} |\la \eta|\gl\ra|^2\gd\(\go-E_{\eta}\)$.
The $A^{\gl_{max}}$ function for the state $\gl_{max}$ corresponding to the
most intense peak of the $\ga$ spectrum is shown in the inset of  Fig.~\ref{fg:azospc}.
In the TDA, the $\gl_{max}$ state is decomposed  only in positive e-h pairs. 
However, the solution the full Hamiltonian reveals an
important contribution from the e-h {\it antipair} with energy $\sim-$2.7\,eV. 
This term, neglected in the TDA, 
causes a dramatic overall redistribution of weights, 
increasing the contribution of e-h pairs at high energies.
Thus, the contribution of the e-h antipair---corresponding to a
spatially extended $\pi^*\rightarrow\pi$ transition---changes the
character of the excitation.  
Whereas in the TDA the most intense excitation in the spectrum can be
identified with a single e-h pair, in the full $H$ solution it acquires a
more collective, plasmon-like character. 

The example of  the {\it trans}-azobenzene makes clear 
that a proper description of the electron-electron correlations in confined systems
requires the solution of the \emph{full} BS/TDDFT Hamiltonians, beyond the TDA. 
However, for larger nanostructures with many degrees of freedom, 
the size  of the $H$ matrix can be as large as $10^6\times 10^6$ and
consequently the problem is
impossible to treat if not using iterative methods.
The TDA reduces $H$ to
a Hermitian Hamiltonian, and makes possible to use the efficient and
stable Hermitian iterative approaches~\cite{Bai00}.   
Therefore we need an iterative approach for calculating the
resolvent of the full non-Hermitian
Hamiltonian $H$ as efficient and stable as for the Hermitian case.

In what follows, we show that it is indeed possible
to design such an iterative approach by observing that
$H$ belongs 
to a class of non-Hermitian Hamiltonians with a real
spectrum\,---that is real eigenvalues. As established by Mostafazadeh~\cite{Mostafazadeh02b}, 
the reality of the spectrum is related to the existence
of a positive-definite inner product with respect to which the Hamiltonian
is Hermitian. 
We show that
for the BS (TDDFT) Hamiltonian---and in general
for all the Hamiltonians of this form---this inner product does
exist (thus the spectrum is real) and, more importantly, it is
\emph{explicitly known}. The knowledge of this product 
allows one to
conveniently transform the iterative approach designed for the
Hermitian TDA Hamiltonian to treat the full non-Hermitian Hamiltonian.   

Following Zimmermann~\cite{Zimmermann70} we can write $H$ as
the product of two noncommuting Hermitian
matrices, 
\begin{equation}\label{eq:efhbar}
 H = F\bar H =
\begin{pmatrix}
1  & 0 \\
0  & -1
\end{pmatrix}
\begin{pmatrix}
R  & C \\
C^{*}  & R^*
\end{pmatrix}.
\end{equation}
One can check that $\bar H H = H^{\dagger} \bar H$. This key
property of $H$ is called $\bar H$-pseudo-Hermicity~\cite{Mostafazadeh02a}.
Through the $\bar H$ operator,
we can define a positive-definite 
$\bar H$-inner product~\cite{note5}
$\langle \cdot | \cdot \rangle_{\bar H} := \langle \cdot
|\bar H| \cdot \rangle$, and a corresponding $\bar H$-expectation value
$\langle \cdot | O| \cdot \rangle_{\bar H} := \langle \cdot
|\bar HO| \cdot \rangle$.
With respect to this  $\bar H$-expectation value,
$H$ is Hermitian as can be verified by using the $\bar H$-pseudo-Hermicity,
\begin{equation}\label{eq:hbrhrm}
\langle v | H |v'  \rangle_{\bar H} = \langle v' | H^{\dagger} \bar
H  |v  \rangle ^* =\langle v' | \bar H H |v
\rangle^* =: \langle v' | H |v  \rangle_{\bar H}^*.
\end{equation}
It follows that the  $\bar H$-expectation value of the resolvent of
$H$,  $(\omega- H)^{-1}$, is Hermitian as well. Then,  
to evaluate $\la P|(\omega- H)^{-1}|P\ra$,  and thus $\epsilon(\omega)$, 
we rewrite it in terms of Hermitian 
$\bar H$-expectation values by using the completeness relationship
$I = \sum_{k} |q_k\rangle \langle q_k |\bar H$
\begin{equation}\label{eq:resrel}
  \langle P |(\omega- H)^{-1} | P \rangle =
\sum_k \langle P |q_k\rangle \langle q_k |(\omega- H)^{-1} | P
\rangle_{\bar H},
\end{equation}
where $\{|q_k\ra\}$ is a complete basis, orthonormal with respect to
  the $\bar H$-inner product. 
The $\langle q_k |(\omega- H)^{-1} | P \rangle_{\bar H}$ are
conveniently calculated within the standard Lanczos--Haydock (LH)
iterative method~\cite{Haydock80}---the same used in the TDA Hermitian
case~\cite{BenedictS99}---\emph{provided that the 
${\bar H}$-inner product replaces the standard one}.
The LH method recursively builds the $\{|q_k\ra\}$ basis 
in which $H$ is represented by
a one-dimensional semi-infinite chain of sites with only 
nearest-neighbors interactions. For such a system the evaluation of
the matrix elements of $(\omega- H)^{-1}$ reduces to the calculation of a
continued fraction.

This approach allows us to treat systems as large as commonly done using the
TDA, but using the full Hamiltonian. 
Computationally, it requires a single matrix--vector
multiplication at each iteration
---as in the Hermitian case~\cite{note7}.  
Since $\langle P |(\omega- H)^{-1} | P \rangle$ in Eq.~(\ref{eq:resrel}) is
converged after a number of iterations
much smaller than the dimension of $H$, this method is far more 
efficient than performing the diagonalization of the Hamiltonian.
Specifically, for the \emph{trans}-azobenzene, 
the number of operations performed
in the diagonalization is about two orders of
magnitude larger than in our approach~\cite{note4},
while the results are 
indistinguishable (Fig.~\ref{fg:azospc}).

\begin{figure}[H]
\begin{center}
\epsfig{figure=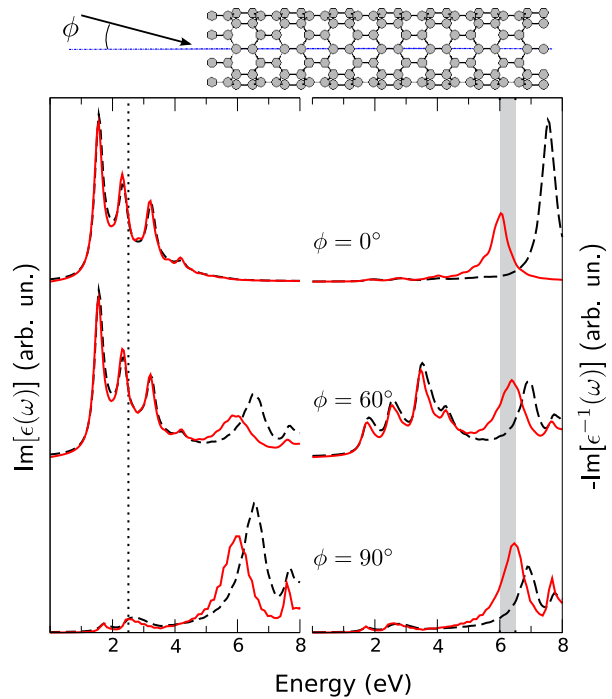,
        clip=,bbllx=50,bblly=30,bburx=585,bbury=645,width=8cm}
\end{center}
\vspace{-.5cm}
\caption{
\footnotesize{
Polarized (angle $\phi$ with respect to the CNT axis) absorption 
(left stack) and $\phi$-dependent EEL (right stack) spectra of
\emph{zig-zag} (8,0) CNT calculated within
the BSE either by using the full (solid
line) or the TDA (dashed line) Hamiltonian. The dotted line indicates the
position of the quasiparticle band gap~\cite{Spataru08}. The gray area shows the experimental energy blueshift (0.5\,eV)
between the parallel and transverse plasmon mode~\cite{Kramberger08},
correctly reproduced by full Hamiltonian calculations. Instead, the TDA
predicts the transverse plasmon mode to red-shift, in  striking
disagreement with the experiment.}}
\label{fig:cntspc} 
\end{figure}

As an application we consider the frequency-dependent response of
carbon nanotubes (CNTs) within the BS equation.  
Recently, experimental studies have shown CNTs to be 
characterized by strongly anisotropic
electronic and optical properties~\cite{Murakami2005,Kramberger08}. 
Moreover,
the measured optical and EEL spectra have revealed that 
excitations
appearing at energies higher than 4\,eV have a collective
character~\cite{Murakami2005,Kramberger08}. 
Using the present method we show that the external field polarization
and the excitonic--plasmonic character of the optical excitations
of CNTs are intimately related. 

So far \emph{ab initio} studies have been limited to the
excitonic effects appearing {\it below} 4\,eV and only 
for longitudinal polarized light~\cite{Spataru08}.
In fact, calculations of the optical response to
transverse perturbing fields are extremely challenging, as  
the plasmonic character of the excitation requires the inclusion
of a huge number of e-h pairs. Using the method proposed in this work
we were able to include up to $\sim 1.4\times 10^{5}$ e-h pairs in the solution of the 
BS equation, {\it without using the TDA}.

Figure~\ref{fig:cntspc} shows the absorption (left stack) and $\qlim$
EEL (right stack) spectrum of CNTs
calculated within the BS~\cite{note1,note3} equation as a function of the angle 
$\phi$ that the perturbing field forms with the tube axis.
We compare the full solution of the BS equation (full line) with the 
result of the TDA (dashed line).

For a parallel perturbing field ($\phi = 0$) the CNT behaves as an
extended solid. 
As expected, in this case the TDA describes well the excitonic peaks
in the absorption
spectrum~\cite{Spataru08}. On the contrary the TDA completely fails in 
reproducing the longitudinal plasmonic peak,
overestimating its frequency by 1.5\,eV. By increasing the $\phi$ angle the 
perturbing field acquires a perpendicular component,  and
a peak at $\sim$\,6\,eV appears in the $\eps\(\go\)$ function. 
The very same peak occurs in the EEL spectrum, confirming that it
corresponds to a mixed
excitonic--plasmonic excitation. This mixed behavior is 
misdescribed by the TDA both in the
absorption and in the EEL spectra. 
The TDA performs even worse 
for a transverse perturbing field 
($\phi=90^\circ$), where the excitations are confined within the tube radius. The
$\eps\(\go\)$ and $\eps^{-1}\(\go\)$ functions become very similar---
that is, the system behaves as an isolated molecule. Thus, like in the case of
the {\it trans}-azobenzene, the contribution from 
e-h antipairs cannot be neglected, and the TDA yields an error on the position
of the main peak in the absorption/EEL spectra of
$\sim$0.6 eV.

Our results for the CNT show that,
in contrast to the common belief,
the excitations appearing in $\Im(\eps)$ are not all purely
excitonic.
When the excitation is forced to be spatially confined, it acquires a 
mixed excitonic--plasmonic character, and the TDA breaks down.
The TDA performs well {\em only} for purely excitonic states that are dominant
only when the perturbing field is polarized along the tube axis.
However, in practice it is not possible to measure selectively only longitudinal
polarization as CNTs are generally randomly oriented in a
sample. Even in the case of vertical aligned CNTs the tubes are found
to form angles of $\sim$25$^\circ$~\cite{Kramberger08}.    

Finally, an important confirmation of the accuracy of the present approach  is
given by the position of the plasmon peaks in the EEL spectra.
Using the full solution of the BS equation, the  plasmon energy 
is found to blue-shift of  $\sim$0.5\,eV  by moving from
$\phi=0^\circ$ to $\phi=90^\circ$.
This is in striking agreement with the $0.5$\,eV value measured in momentum
dependent EEL experiments~\cite{Kramberger08}. 
On the contrary, the TDA causes a 
energy {\it redshift} of the plasmon, in disagreement
with the experimental results.

In summary, we have shown that the TDA  breaks down
in nanoscale systems where dimensionality effects confine
the optical excitation, inducing a 
mixed excitonic--plasmonic behavior.
We propose a novel approach to solve the BS/TDDFT equations
beyond the TDA, keeping the numerical advantages of 
a Hermitian formulation.  
This approach successfully explains the experimental features
in the optical and EEL spectra of CNTs,
and opens the way to a truly \ai approach to linear response properties
of nanoscale materials.

\begin{acknowledgments}
This work was supported by the EU through the FP6 Nanoquanta NoE
(NMP4-CT-2004-50019), the FP7 ETSF I3 e-Infrastructure (Grant Agreement
211956), the Belgian Interuniversity Attraction Poles Program (P6/42), the
Communaut\'e Fran\c{c}aise de Belgique (ARC 07/12-003), the R\'egion Wallone (WALL-ETSF).

\end{acknowledgments}

\bibliographystyle{apsrev}

\begin{thebibliography}{22}
\expandafter\ifx\csname natexlab\endcsname\relax\def\natexlab#1{#1}\fi
\expandafter\ifx\csname bibnamefont\endcsname\relax
  \def\bibnamefont#1{#1}\fi
\expandafter\ifx\csname bibfnamefont\endcsname\relax
  \def\bibfnamefont#1{#1}\fi
\expandafter\ifx\csname citenamefont\endcsname\relax
  \def\citenamefont#1{#1}\fi
\expandafter\ifx\csname url\endcsname\relax
  \def\url#1{\texttt{#1}}\fi
\expandafter\ifx\csname urlprefix\endcsname\relax\def\urlprefix{URL }\fi
\providecommand{\bibinfo}[2]{#2}
\providecommand{\eprint}[2][]{\url{#2}}

\bibitem[{\citenamefont{Salpeter and Bethe}(1951)}]{SalpeterB51}
\bibinfo{author}{\bibfnamefont{E.~E.} \bibnamefont{Salpeter}} \bibnamefont{and}
  \bibinfo{author}{\bibfnamefont{H.~A.} \bibnamefont{Bethe}},
  \bibinfo{journal}{Phys. Rev.} \textbf{\bibinfo{volume}{84}},
  \bibinfo{pages}{1232} (\bibinfo{year}{1951}).

\bibitem[{\citenamefont{Runge and Gross}(1984)}]{RungeG84}
\bibinfo{author}{\bibfnamefont{E.}~\bibnamefont{Runge}} \bibnamefont{and}
  \bibinfo{author}{\bibfnamefont{E.~K.~U.} \bibnamefont{Gross}},
  \bibinfo{journal}{Phys. Rev. Lett.} \textbf{\bibinfo{volume}{52}},
  \bibinfo{pages}{997} (\bibinfo{year}{1984}).

\bibitem[{\citenamefont{Fetter and Walecka}(2003)}]{FetterW71}
\bibinfo{author}{\bibfnamefont{A.~L.} \bibnamefont{Fetter}} \bibnamefont{and}
  \bibinfo{author}{\bibfnamefont{J.~D.} \bibnamefont{Walecka}},
  \emph{\bibinfo{title}{Quantum Theory of many-particle systems}}
  (\bibinfo{publisher}{Dover}, \bibinfo{year}{2003}),
  chap.~\bibinfo{chapter}{15}, p. \bibinfo{pages}{565}.

\bibitem[{\citenamefont{Bai et al}(2000)}]{Bai00}
\emph{\bibinfo{booktitle}{Templates for the solution of algebraic
    problems: a practical guide}}, edited by
\bibinfo{editor}{\bibfnamefont{Z.} \bibnamefont{Bai}},
\bibinfo{editor}{\bibfnamefont{J.} \bibnamefont{Demmel}},
\bibinfo{editor}{\bibfnamefont{J.} \bibnamefont{Dongarra}},
\bibinfo{editor}{\bibfnamefont{A.} \bibnamefont{Ruhe}},
\bibinfo{editor}{\bibfnamefont{H.} \bibnamefont{van der Vorst}},
   (\bibinfo{publisher}{SIAM}, \bibinfo{address}{Philadelphia},
   \bibinfo{year}{2000}).

\bibitem{OlevanoR01} V. Olevano and L. Reining, Phys. Rev. Lett. {\bf 86},
  5962 (2001).

\bibitem{More}
See e.g. 
M. Del Puerto et al. Phys. Rev. Lett. {\bf 97}, 096401 (2006);
B. Arnaud et al. \bibinfo{journal}{\textit{ibid.}} {\bf 96}, 026402 (2006);
L. Wirtz et al., \bibinfo{journal}{\textit{ibid.}}, 126104 (2006)
\bibitem{Spataru08} C.D. Spataru et al., Top. Appl. Phys. {\bf 111}, 195 (2008).



\bibitem{Hirata99} S. Hirata and M. Head-Gordon,
  Chem. Phys. Lett. {\bf 314}, 291 (1999).

\bibitem[{\citenamefont{Onida et~al.}(2002)\citenamefont{Onida, Reining, and
  Rubio}}]{OnidaRR02}
\bibinfo{author}{\bibfnamefont{G.}~\bibnamefont{Onida}},
\bibnamefont{et al.},
  \bibinfo{journal}{Rev. Modern Phys.} \textbf{\bibinfo{volume}{74}},
  \bibinfo{pages}{601} (\bibinfo{year}{2002}).

\bibitem{note1} BS/TDDFT spectra are
  calculated with the {\sc yambo} code,
  (A. Marini et al., the {\sc yambo} project, http://www.yambo-code.org), where the
  algorithm proposed in this work has been implemented.
  The Kohn--Sham basis (local density 
  approximation) is calculated with {\sc abinit},
   X. Gonze et al. Comp. Mat. Science {\bf 25}, 478 (2002);
  Z. Kristallogr. {\bf 220}, 478 (2005).

\bibitem{note2} We have included e-h pairs and antipairs in a
window of 30\,eV and used a cut-off of 1.6 Ha for
the $\bar V$ integrals. 

\bibitem[{\citenamefont{Mostafazadeh}(2002{\natexlab{a}})}]{Mostafazadeh02b}
\bibinfo{author}{\bibfnamefont{A.}~\bibnamefont{Mostafazadeh}},
  \bibinfo{journal}{J. Math. Phys.} \textbf{\bibinfo{volume}{43}},
  \bibinfo{pages}{3944} (\bibinfo{year}{2002}{\natexlab{a}}).

\bibitem[{\citenamefont{Zimmermann}(1970)}]{Zimmermann70}
\bibinfo{author}{\bibfnamefont{R.}~\bibnamefont{Zimmermann}},
  \bibinfo{journal}{Phys. Stat. Sol.} \textbf{\bibinfo{volume}{41}},
  \bibinfo{pages}{23} (\bibinfo{year}{1970}).

\bibitem[{\citenamefont{Mostafazadeh}(2002{\natexlab{b}})}]{Mostafazadeh02a}
\bibinfo{author}{\bibfnamefont{A.}~\bibnamefont{Mostafazadeh}},
  \bibinfo{journal}{J. Math. Phys.} \textbf{\bibinfo{volume}{43}},
  \bibinfo{pages}{205} (\bibinfo{year}{2002}{\natexlab{b}}).

\bibitem{note5} When the formation
of excitons does not lead to a fundamental change of
ground state, $\bar H$ is positive
definite, see Ref.~\cite{Zimmermann70}.

\bibitem[{\citenamefont{Haydock}(1980)}]{Haydock80}
\bibinfo{author}{\bibfnamefont{R.}~\bibnamefont{Haydock}}, in
  \emph{\bibinfo{booktitle}{Solid State Phys.}},
  edited by \bibinfo{editor}{\bibfnamefont{H.}
    \bibnamefont{Ehrenfest}}, \bibinfo{editor}{\bibfnamefont{F.}
    \bibnamefont{Seitz}},\bibnamefont{and} \bibinfo{editor}{\bibfnamefont{D.}
    \bibnamefont{Turnbull}} 
  (\bibinfo{publisher}{Academic Press}, 
  \bibinfo{year}{1980}), vol.~\bibinfo{volume}{35}, p. \bibinfo{pages}{215--294}.

\bibitem{BenedictS99}
L. X. Benedict and E. L. Shirley, Phys. Rev. B {\bf 59}, 5441 (1999).



\bibitem{note7} Because of the interplay between
  $H$, $F$ and $\bar H$, evaluating the $\bar H$-inner product does not
  require extra operation with respect to the standard inner-product.

\bibitem{note4} 
The proposed approach requires  $mN^2$ operations, where $m$ is the number of
iterations and $N$ the size of the system.  For azobenzene $N\sim10^4$
and $m\sim10^2$. Thus it is much faster and convenient compared to the
standard diagonalization, that requires $O(N^3)$
operations.

\bibitem{note3} We have included e-h pairs and antipairs from band
  32 to 96 in a window of 15 eV and used cut-offs of 4 Ha and 1 Ha for
the $\bar V$ and $W$ integrals. The 1D Brillouin zone
is sampled by 64 k-points. Quasiparticle correction from
  Ref.~\cite{Spataru08}. 

\bibitem[{\citenamefont{Murakami et~al.}(2005)\citenamefont{Murakami,
  Einarsson, Edamura, and Maruyama}}]{Murakami2005}
\bibinfo{author}{\bibfnamefont{Y.}~\bibnamefont{Murakami}},
\bibnamefont{et al.},
  \bibinfo{journal}{Phys. Rev. Lett.} \textbf{\bibinfo{volume}{94}},
  \bibinfo{pages}{087402} (\bibinfo{year}{2005}).

\bibitem[{\citenamefont{Kramberger}(2001)}]{Kramberger08}
\bibinfo{author}{\bibfnamefont{C.}~\bibnamefont{Kramberger}},
\bibnamefont{et al.},
  \bibinfo{journal}{Phys. Rev. Lett.} \textbf{\bibinfo{volume}{100}},
  \bibinfo{pages}{196803} (\bibinfo{year}{2008}).

\end{thebibliography}

\end{document}